# Steganography Based on Pixel Intensity Value Decomposition

Alan Anwer Abdulla, Harin Sellahewa and Sabah A. Jassim
University of Buckingham, Buckingham MK18 1EG, UK
alananwer@yahoo.com {sabah.jassim; harin.sellahewa}@buckingham.ac.uk

## ABSTRACT

This paper focuses on steganography based on pixel intensity value decomposition. A number of existing schemes such as binary, Fibonacci, Prime, Natural, Lucas, and Catalan-Fibonacci (CF) are evaluated in terms of payload capacity and stego quality. A new technique based on a specific representation is used to decompose pixel intensity values into 16 (virtual) bit-planes suitable for embedding purposes. The new decomposition scheme has a desirable property whereby the sum of all bit-planes does not exceed the maximum pixel intensity value, i.e. 255. Experimental results demonstrate that the new decomposition scheme offers a better compromise between payload capacity and stego quality than other existing decomposition schemes used for embedding messages. However, embedding in the 6$^{th}$ bit-plane onwards, the proposed scheme offers better stego quality. In general, the new decomposition technique has less effect in terms of quality on pixel value when compared to most existing pixel intensity value decomposition techniques when embedding messages in higher bit-planes.

## 1. INTRODUCTION

The aim of steganography systems is to communicate a secret message in a way that would not be noticeable by an intruder. It is generally accepted that any steganographic technique must possess two main properties: good stego quality, and sufficient data capacity [1]. The media that is carrying the secret message is called the cover or the carrier. After the secret message is embedded into the cover, the resulting artifact is called the stego. For the data hiding, the cover can be any of the following digital media: image, audio, and video. In this paper, images are used as covers because images usually have a high degree of redundancy, which makes them suitable to embed information without degrading their visual quality.

There are two domains that are regularly used for data hiding, the frequency domain [2-4] and the spatial domain [5,6]. In the frequency domain data hiding techniques, the secret data bits are inserted into the coefficients of the image pixel's frequency representation. Among the frequency image pixel's frequency representation are Discrete Cosine Transform (DCT) [2], Discrete Fourier Transform (DFT) [3] and Discrete Wavelet Transform (DWT) [4]. On the other hand, in the spatial domain techniques, the secret data bits are inserted directly into the images' pixel value decomposition. Replacing the secret bits with the Least Significant Bit (LSB) is considered the most widely used spatial domain approach for data hiding [5, 6]. One well-known bit-plane representation of pixel value is the binary bit-plane and almost all the steganographic methods use binary representation.

In this paper, existing pixel value representations such as: Fibonacci [7, 8], Prime [9], Natural [10], Catalan-Fibonacci (CF) [11], and Lucas [12] are studied and evaluated. Moreover, a new pixel value representation is evaluated.

The underlying idea behind pixel value decomposition techniques according to different number sequences is the partitioning of a pixel intensity value into a higher number of bit-planes than the traditional binary representation with an aim to increase the number of bit-planes suitable for embedding. Intensity values of typical grey scale images range from 0 to 255. These intensity values require 8 bits to represent them in binary, whereas Fibonacci, Prime, Natural, CF, Lucas representation require 12, 15, 23, 15 and12 bits respectively. Each decomposition technique has its advantages and disadvantages. In this contribution, the advantages and disadvantages of existing decomposition scheme have been addressed by comparing them in term of payload capacity and stego quality, and a new decomposition has been tested which can make a balance between these existing decomposition techniques in term of payload capacity and stego quality.

The rest of the paper is organized as follows. A review of literature on relevant schemes is presented in Section 2. The new pixel value decomposition and embedding procedure is presented in Section 3. Finally, experimental results and conclusion are shown in Section 4 and 5 respectively.

## 2. LITERATURE REVIEW

Different pixel value decomposition techniques are proposed to achieve steganography requirements. These decomposition techniques aim to provide more bit-planes than the traditional binary bit-planes to embed the secrets with less effect on stego-quality when secret bits are embedded in higher bit-planes. However, their limitation is payload capacity. Here we review a number of existing pixel value decomposition techniques used in steganography.

In steganography, the most well-known and widely used pixel value decomposition technique is binary. In such decomposition, the LSBs are replaced with the secret bits either in sequential order [5] or randomly [6]. Embedding techniques based on binary decomposition have the advantage of payload capacity. However, embedding in higher bit-planes leads to noticeable distortions in the stego image. To address this, Picione et al. [7] proposed the first decomposition technique over binary decomposition called Fibonacci decomposition which decomposes a pixel value into 12 bit-planes. In [7], the authors claimed that the embedding technique based on the Fibonacci decomposition has a property of increasing bit-planes enabling the embedding in higher bit-planes with less stego quality distortion compared to the binary based embedding. However, the traditional binary representation of a pixel value is unique, whilst the Fibonacci representation is not, i.e., more than one bit stream can represent the same number. A unique Fibonacci representation is obtained by applying Zeckendorf's theorem. For example, the number 5 can be coded as 1000 or 0110. According to the Zeckendorf code, 0110 is not valid. Therefore, embedding based on Fibonacci decomposition technique has the limitation of payload capacity because not every cover pixel is a good candidate for embedding.

Following Fibonacci decomposition, Prime decomposition was proposed by Dey et al. [9] where each pixel value is decomposed into 15 bit-planes. The authors claimed that an improvement over binary and Fibonacci decomposition has been made since the number of bit-planes has increased to 15. They also claimed that their proposed decomposition technique not only allows one to embed a secret message in higher bit-planes but do it without much distortion, i.e., achieve much better stego quality compared to Fibonacci and binary decompositions. However, the limitation of this technique is payload capacity since not every cover pixel is usable for embedding.

In [10], a new pixel value decomposition called Natural decomposition was proposed. Each pixel value is, lexicographically greatest, decomposed into 23 bit-planes. The authors claimed that this decomposition is done as an improvement over binary, Fibonacci, and Prime decomposition since the number of bit-planes has increased to 23. They also claimed that the proposed embedding technique not only allows one to embed secret messages in higher bit-planes but it can be done with less distortion. However, the limitation of this technique is payload capacity (has the minimum capacity out of all other decomposition techniques studied in this paper; see Figures 2-9), since not every cover pixel can be used for embedding. The next pixel value decomposition, proposed by Aroukatos et al., is called Catalan-Fibonacci (CF) [11]. This decomposition is based on a sequence of numbers that are obtained by the union of a sub set of Fibonacci numbers and sub set of Catalan numbers. Relying on this particular sequence of numbers, each pixel value is decomposed, lexicographically greatest, into 15 bit-planes. This decomposition is proposed as an improvement over Fibonacci based embedding, since using larger set of bit-planes provides more bit-planes that can be used for embedding. The authors claimed that this proposed embedding technique is allows one to embed secret message in higher bit-planes and that this can be achieved without much distortion. The limitation of this technique is its payload capacity, since not every cover pixel can be used for embedding.

Finally, another pixel value decomposition based on Lucas numbers is proposed by Alharbi [12]. This decomposition technique decomposes a pixel value into 12 bit-planes aimed to allow using higher bit-plane for embedding without much degrading the stego quality comparing to binary. The author claimed that this decomposition is done as an improvement over binary, since using larger set of virtual bit-planes provides more bit-planes that can be used for embedding. The limitation of this technique is stego quality, because embedding a secret bit in LSB could change the cover pixel value by 2 – the first bit-plane of the Lucas has a value of 2. There is another limitation of Lucas which is payload capacity because not every cover pixel is usable for embedding. The new pixel value decomposition technique is presented in the next section.

## 3. NEW PIXEL VALUE DECOMPOSITION TECHNIQUE AND EMBEDDING PROCEDURE

In this section, both the new pixel value decomposition technique and the embedding procedure based on it are discussed.

### 3.1 New pixel value decomposition technique

The new pixel value decomposition technique is based on a set of numbers that can be used to embed data in higher bit-planes without incurring much degradation on the stego image quality. The set of numbers S can be defined as:

$$S = \{1\} \cup \{2n \mid 1 \leq n \leq 16, \text{ where } n \neq 9\} \quad (1)$$

In other words, S = {1, 2, 4, 6, 8, 10, 12, 14, 16, 20, 22, 24, 26, 28, 30, 32}. The reason for excluding number 18 is that the summation of the set S must equal to the 255. The new decomposition technique has the following properties:

1. The summation of the numbers results in value 255, the upper bound of the 8-bits digital boundary.
2. All natural numbers between 0-255 can be represented.

According to the set S, each pixel value $P$ is decomposed into 16 bits and the weight of the bit-plane can be defined as:

$$P = \sum_{i=1}^{16} b_i W_i \quad (2)$$

$$\text{where } b_i \in \{0,1\} \text{ and } W_i = \begin{cases} 1 & \text{if } i = 1 \\ 2(i-1) & \text{if } 2 \leq i \leq 9 \\ 2i & \text{if } 10 \leq i \leq 16 \end{cases} \quad (3)$$

If any pixel value has more than one representation in this number system, the lexicographically highest of them is always taken, to assert invertible property (e.g., the number 12 has two different representation in 6 bits of the new number system, namely 100010 and 010100 since there are:

(1 * 10) + (0 * 8) + (0 * 6) + (0 * 4) + (1 * 2) + (0 * 1) = 12 and (0 * 10) + (1 * 8) + (0 * 6) + (1 * 4) + (0 * 2) + (0 * 1) = 12

As 100010 lexicographically (from left to right) is higher than 010100, 100010 will be chose to be the valid representation for number 12 in the new number system and thus we discard 010100 as an invalid representation. Therefore 12 ≡ max$_{lexicographic}$ (100010, 010100) ≡ 100010, see Table 1 for the valid representation of numbers from 0 to 44. Table 2 illustrates the number of bit-planes and their weight for the existing systems of pixel value decomposition and the new one.

### 3.2 Embedding procedure

For the embedding procedure, first the selected cover pixel value (pseudorandom number generator (PRNG) is used to randomly select a cover pixel) is converted using the new decomposition technique into 16 bits. Now the secret bit can be embedded into an agreed (virtual) bit-plane by simply replacing the corresponding bit, only if we find that after embedding the resulting representation is valid representation in the new number system, otherwise do not embed and skip. This is only to guarantee the extraction of the inverse and correctness for extraction of the secret embedded message. After embedding the secret message, we convert the resultant sequence in the new number system back to its value (decimal value) and get the stego-image. The extraction procedure is exactly the reverse. From stego-image, we convert each selected pixel, using the same PRNG used at the embedding stage; with embedded data bit to its corresponding new decomposition and from the agreed bit-plane extract the secret bit. Combine all bits to get the secret message.

Table 1: Decimal number and its decomposition using the new decomposition system

| N | Proposed decomp. | N | Proposed decomp. | N | Proposed decomp. |
|---|---|---|---|---|---|
| 0 | 0000000000000000 | 15 | 0000000010000001 | 30 | 0100000000000000 |
| 1 | 0000000000000001 | 16 | 0000000100000000 | 31 | 0100000000000001 |
| 2 | 0000000000000010 | 17 | 0000000100000001 | 32 | 1000000000000000 |
| 3 | 0000000000000011 | 18 | 0000000100000010 | 33 | 1000000000000001 |
| 4 | 0000000000000100 | 19 | 0000000100000011 | 34 | 1000000000000010 |
| 5 | 0000000000000101 | 20 | 0000001000000000 | 35 | 1000000000000011 |
| 6 | 0000000000001000 | 21 | 0000001000000001 | 36 | 1000000000000100 |
| 7 | 0000000000001001 | 22 | 0000010000000000 | 37 | 1000000000000101 |
| 8 | 0000000000010000 | 23 | 0000010000000001 | 38 | 1000000000001000 |
| 9 | 0000000000010001 | 24 | 0000100000000000 | 39 | 1000000000001001 |
| 10 | 0000000000100000 | 25 | 0000100000000001 | 40 | 1000000000010000 |
| 11 | 0000000000100001 | 26 | 0001000000000000 | 41 | 1000000000010001 |
| 12 | 0000000001000000 | 27 | 0001000000000001 | 42 | 1000000000100000 |
| 13 | 0000000001000001 | 28 | 0010000000000000 | 43 | 1000000000100001 |
| 14 | 0000000010000000 | 29 | 0010000000000001 | 44 | 1000000001000000 |

The first, third, and fifth columns of the Table 1 are the decimal numbers from 0 to 44 (consider them as pixel values), and the second, fourth, and sixth columns are decomposing these values into 16 bits using the new decomposition system.

Table 2: Number of bit-planes and its corresponding weight for different pixel value decomposition systems

| L | Binary | Fibonacci | Lucas | Prime | CF | New technique | Natural |
|---|---|---|---|---|---|---|---|
| 1 | 1 | 1 | 2 | 1 | 1 | 1 | 1 |
| 2 | 2 | 2 | 1 | 2 | 2 | 2 | 2 |
| 3 | 4 | 3 | 3 | 3 | 3 | 4 | 3 |
| 4 | 8 | 5 | 4 | 5 | 5 | 6 | 4 |
| 5 | 16 | 8 | 7 | 7 | 8 | 8 | 5 |
| 6 | 32 | 13 | 11 | 11 | 13 | 10 | 6 |
| 7 | 64 | 21 | 18 | 13 | 14 | 12 | 7 |
| 8 | 128 | 34 | 29 | 17 | 21 | 14 | 8 |
| 9 | | 55 | 47 | 19 | 34 | 16 | 9 |
| 10 | | 89 | 76 | 23 | 42 | 20 | 10 |
| 11 | | 144 | 123 | 29 | 55 | 22 | 11 |
| 12 | | 233 | 199 | 31 | 89 | 24 | 12 |
| 13 | | | | 37 | 132 | 26 | 13 |
| 14 | | | | 41 | 144 | 28 | 14 |
| 15 | | | | 43 | 233 | 30 | 15 |
| 16 | | | | | | 32 | 16 |
| 17 | | | | | | | 17 |
| 18 | | | | | | | 18 |
| 19 | | | | | | | 19 |
| 20 | | | | | | | 20 |
| 21 | | | | | | | 21 |
| 22 | | | | | | | 22 |
| 23 | | | | | | | 23 |

From Table 2, first column, L, is the number of bit-planes for each pixel value decomposition system. Each cell is representing the weight value from 1st LSB to ith MSB (from top to bottom) for the corresponding bit-plane of each decomposition system. Table 2 is showed that by embedding the secret bits from 6th LSB to higher ith MSB, the new technique has minimum stego degradation than all other decomposition systems except Natural.

## 4. EXPERIMENTAL RESULTS

Two experiments are conducted to evaluate the new decomposition technique: first experiment to evaluate payload capacity, and the second experiment to evaluate stego quality. The results are compared with the existing pixel value decomposition systems used for embedding [7, 9, 10, 11, 12]. We used 10 cover-images of size 512 x 512 (see Figure 1) and the secret bits have been generated using a PRNG. After embedding, for each case, i.e. for each technique, 10 stegos are produced.

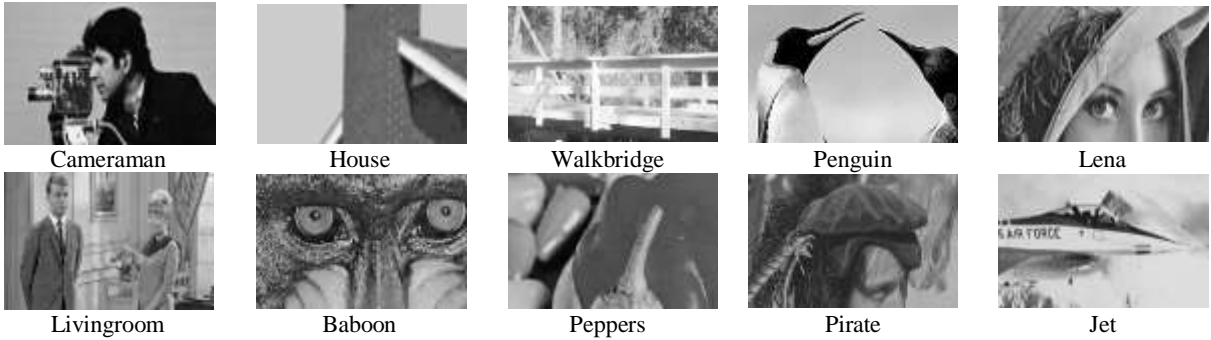

Figure 1. Test images

### 4.1 Payload capacity test

To evaluate the payload capacity, for each technique, bit-planes from 1st LSB to 8th MSB have been tested as presented in Figures (2 – 9).

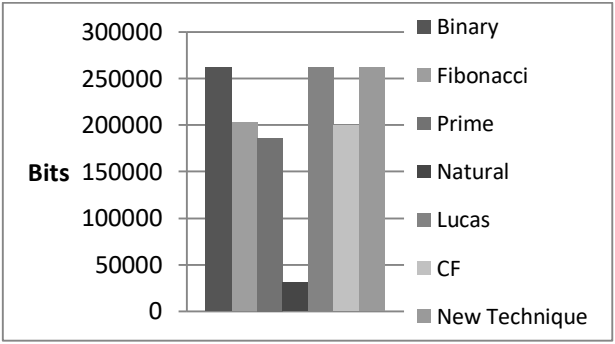

Figure 2. Payload capacity of 1st LSB

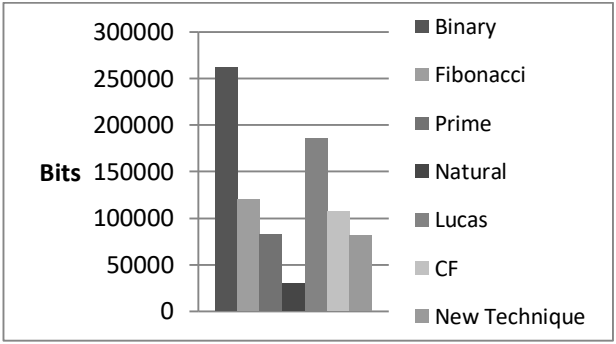

Figure 3. Payload capacity of 2nd LSB

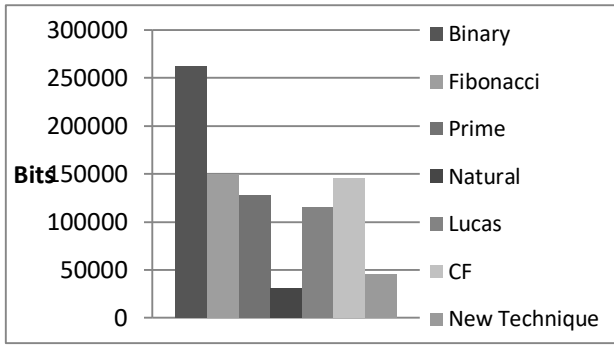

Figure 4. Payload capacity of 3rd LSB

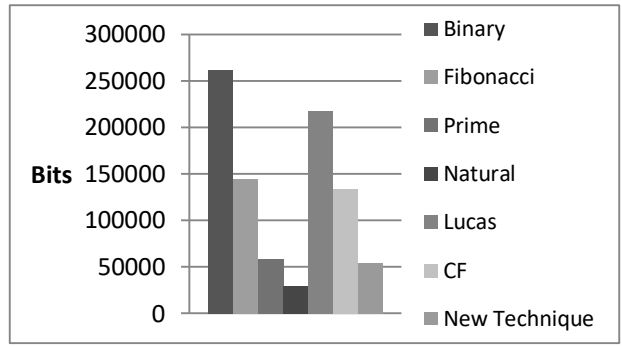

Figure 5. Payload capacity of 4th LSB

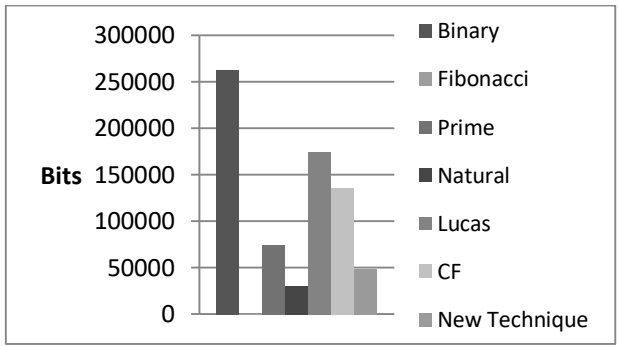

Figure 6. Payload capacity of 5th LSB

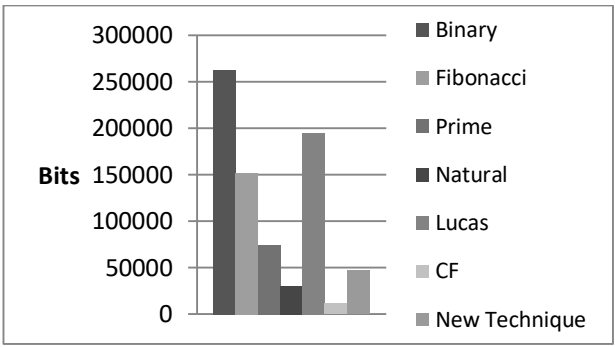

Figure 7. Payload capacity of 6th LSB

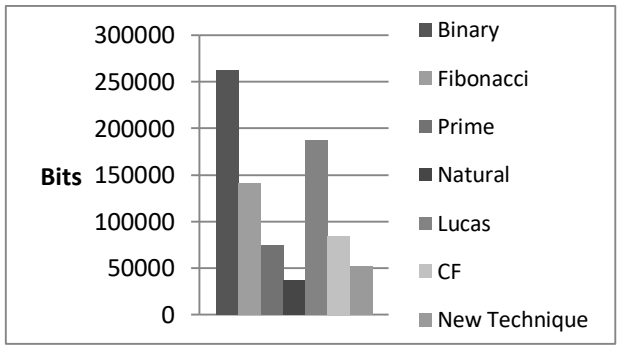

Figure 8. Payload capacity of 7th LSB

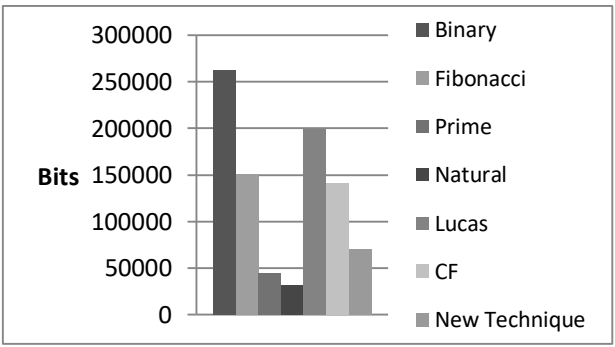

Figure 9. Payload capacity of 8th MSB

From Figure 2, it is noticeable that binary, Lucas, and the new number system based embedding technique have a maximum capacity because every cover pixel is usable for embedding when the secret bits are embedded in 1st LSB, and in Figures (2-9) the Natural embedding based has lowest capacity, 31000 bits across, because not every cover pixel is usable for embedding and depending on the nature of the Natural sequence, after embedding the secret bit, the sequence of bits are changed. For example, if the pixel value is 3 (i.e. 100 in Natural code) and the secret bit is 1, then by replacing the secret bit with first bit-plane we get (101) which is equal to number 4 in decimal. Then by converting number 4 into Natural decomposition we get (1000). Now by extracting from first bit-plane, we get 0 instead 1. Therefore such a pixels value are skipped and not used for embedding the secrets. In Figures (2-9), the new number system based embedding has higher capacity than the Natural based embedding. From Figures 3, and 9, one can notice that the new number system based embedding has higher capacity than the Prime based embedding, and in Figure 7 one can see that the new number system based embedding has higher capacity than the CF based embedding. Moreover, the new number system based embedding has higher capacity than Fibonacci when the secret bits are embedded in the 5th LSB, see Figure 6.

## 4.2 Stego-quality test

To evaluate the stego-quality, Peak signal-to-noise ratio (PSNR) has been used as a measure of the quality. A larger PSNR value means that the stego image preserves the original cover image quality better. Figures (10-17) are illustrated the average of PSNR for 10 stego-images by embedding the same secret bits in each bit-plane from $1^{st}$ LSB to $8^{th}$ MSB separately.

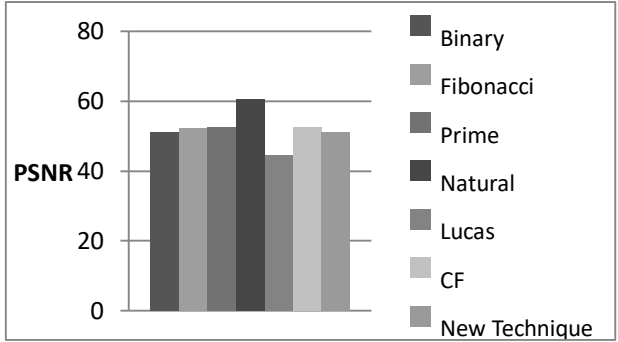

Figure 10. Stego quality by embedding in $1^{st}$ LSB

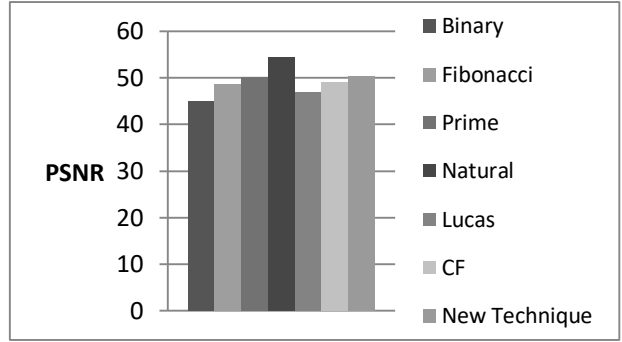

Figure 11. Stego quality by embedding in $2^{nd}$ LSB

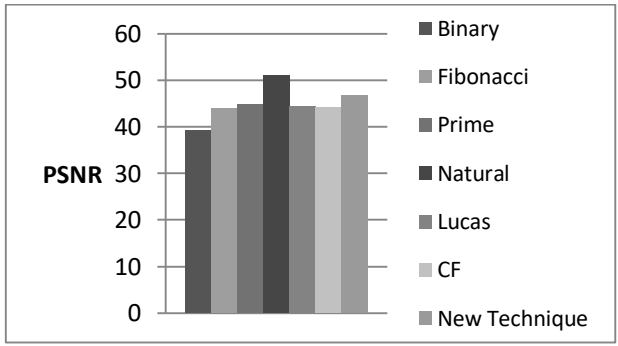

Figure 12. Stego quality by embedding in $3^{rd}$ LSB

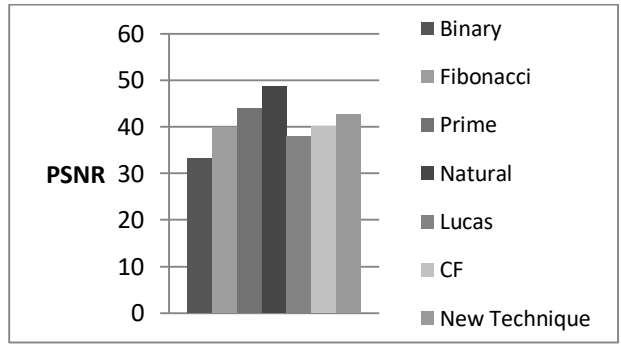

Figure 13. Stego quality by embedding in $4^{th}$ LSB

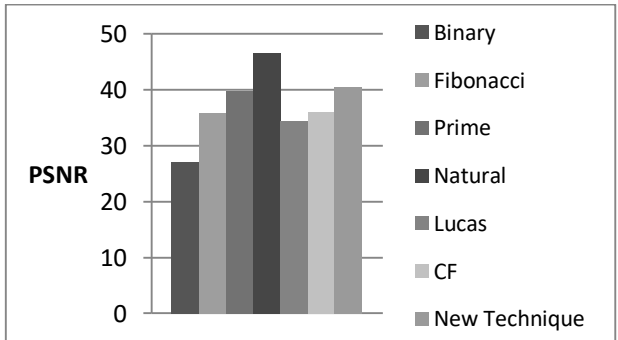

Figure 14. Stego quality by embedding in $5^{th}$ LSB

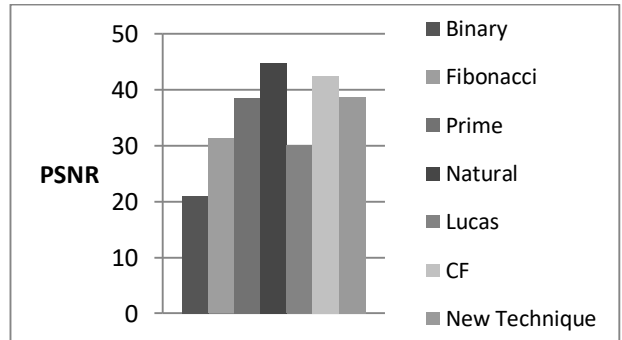

Figure 15. Stego quality by embedding in $6^{th}$ LSB

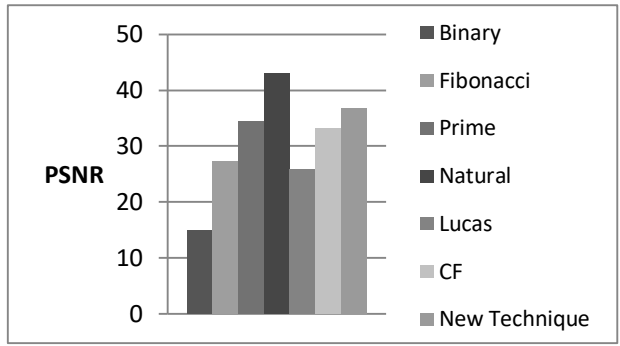
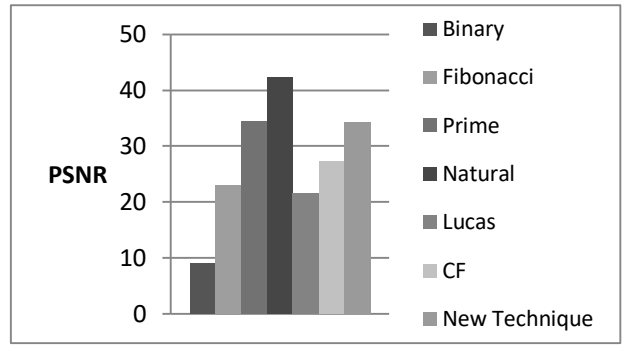

Figure 16. Stego quality by embedding in 7$^{th}$ LSB

Figure 17. Stego quality by embedding in 8$^{th}$ MSB

From Figures 10 – 17, one can notice that the PSNR of Natural based embedding is higher than all others, this is because Natural is able to carry the minimum number of secret bits (see Figures 2-9). In the Figure 10, although binary, Lucas, and the new technique has the same capacity when the secret bits are embedded in 1$^{st}$ LSB (see Figure 2), but the quality of Lucas is lower than the binary and new technique, this is happened because the weight of 1$^{st}$ LSB of Lucas is 2 (see Table 2). From Figures 11 and 17, it is noticeable that in the case of embedding secret bits in 2$^{nd}$ LSB or 8$^{th}$ MSB, although the new technique has higher capacity than the Prime (see Figures 3 and 9), and also has higher quality/ PSNR, i.e. the new technique can carries higher number of secret bits than the Prime based and also has higher PSNR. While embedding based on the new technique decomposition system has higher capacity than Fibonacci when the secret bits are embedded in 5$^{th}$ LSB, see Figure 6, it also has higher PSNR than Fibonacci based, see Figure 14.

### 4.3 Comparison between existing pixel value decomposition techniques

In order to make the comparison between existing pixel value decomposition systems including the new number system, Table 3 illustrates the advantages and disadvantages of these systems:

Table 3: Comparison between decomposition systems

| | Sequence of numbers | Bits | Every Pixel used for 1$^{st}$ LSB | Every Pixel used for other than 1$^{st}$ LSB | Quality distortion |
|---|---|---|---|---|---|
| B | 1,2,4,8,16,32,64,128 | 8 | Yes | Yes | It has maximum distortion compared to all other existing decomposition systems. |
| F | 1,2,3,5,8,13,21,34,55,89,144,233 | 12 | No | No | Less distortion than binary system. |
| Pr | 1,2, 3, 5, 7, 11, 13, 17, 19, 23, 29, 31, 37, 41, 43 | 15 | No | No | Has less distortion than binary, Fibonacci, Lucas, and Catalan |
| N | 1,2,3,4,5,6,7,8,9,10,11,12,13,14,15,16,17,18,19, 20,21,22,23 | 23 | No | No | Has less distortion than all other decomposition systems |
| L | 1, 2, 3, 4, 7, 11, 18, 29, 47, 76, 123, 199 | 12 | Yes | No | Has less distortion than binary and Fibonacci |
| C | 1,2,3,5,8,13,14,21, 34, 42, 55, 89, 132, 144, 233 | 15 | No | No | Has less distortion than binary, Fibonacci, and Lucas |
| Ns | 1,2,4,6,8,10,12,14, 16, 20, 22, 24, 26, 28, 30, 32 | 16 | Yes | No | Has less distortion than all other systems except Natural system |

B, F, Pr, N, L, C and Ns in the first column of Table 3 are referring to binary, Fibonacci, Prime, Natural, Lucas, Catalan-Fibonacci, and the new decomposition systems respectively. The second column is the sequence of numbers of each decomposition system, and the third column is the number of bit-planes. Fourth column is meaning that whether every cover pixel is used for embedding or not when the secret bits are embedded in 1$^{st}$ LSB, and fifth column refers to whether every cover pixel is used for embedding when the secrets are embedded in higher bit-planes, other than 1$^{st}$ LSB. Sixth column is the comparison between the decomposition systems in term of stego quality distortion when the secret bits are embedded in higher bit-planes.

## 5. CONCLUSION

Existing systems for pixel value decomposition such as binary, Fibonacci, Prime, Natural, Lucas, and Catalan-Fibonacci (CF) that are used for data hiding have been studied and evaluated in terms of payload capacity and stego quality. Also a new technique based on a specific representation is used to decompose pixel intensity values into 16 (virtual) bit-planes suitable for embedding purposes. Experimental results demonstrate that the new technique offers a better compromise between payload capacity and stego quality than other decomposition schemes used for embedding secrets. In general, the new decomposition scheme has less effect on pixel value when compared to most existing pixel intensity value decomposition techniques when embedding messages in higher bit-planes. The limitation of the new decomposition scheme is payload capacity when the secret bits are embedded in higher bit-planes because not every pixel is usable for embedding and this is the case in all other decomposition systems except binary based embedding.

## REFERENCES

[1] Thomas, P., "Literature Survey on Modern Image Steganographic Techniques," International Journal of Engineering 5(2), (2013).

[2] Kekre, HB, Athawale, A.A. and Pallavi, N.H.," Increased Capacity and High Security for Embedding Secret Message in Transform Domain using Discrete Cosine Transform," Sci. Engg. & Tech. Mgt. 2 (2), (2010).

[3] Ghoshal, N., and Mandal, J., "A novel technique for image authentication in frequency domain using discrete Fourier transformation technique (IAFDDFTT)," Malaysian Journal of Computer Science 21(1), 24-32, (2008).

[4] Yang, B., and Deng, B., "Steganography in gray images using wavelet," Proc. ISCCSP, (2006).

[5] Chan, C., and Cheng, L. M.,"Hiding data in images by simple LSB substitution," Pattern recognition 37(3), 469-474 (2004).

[6] Thien, C. C., and Lin, J.,"A simple and high-hiding capacity method for hiding digit-by-digit data in images based on modulus function," Pattern Recognition 36(12), 2875-2881 (2003).

[7] Picione, D. D.L., Battisti, F., Carli, M., Astola, J., and Egiazarian, K.,"A Fibonacci LSB data hiding technique," Proc. EUSIPCO, (2006).

[8] Battisti, F., Carli, M., Neri, A., and Egiaziarian, K., "A Generalized Fibonacci LSB Data Hiding Technique," Proc. CODEC, (2006).

[9] Dey, S., Abraham, A., and Sanyal, S., "An LSB Data Hiding Technique Using Prime Numbers," Proc. Information Assurance and Security (IAS) IEEE, 101-108 (2007).

[10] Dey, S., Abraham, A., and Sanyal, S., "An LSB Data Hiding Technique Using Natural Number Decomposition," Proc. IIHMSP, IEEE, 473-476 (2007).

[11] Aroukatos, N., Manes, K., Zimeras, S., and Georgiakodis, F., "Data Hiding Techniques in Steganography using Fibonacci and Catalan numbers," Proc. Information Technology: New Generations (ITNG), IEEE, 392-396 (2012).

[12] Alharbi, F.,"Novel Steganography System using Lucas Sequence," International Journal of Advanced Computer Science and Applications (IJACSA) 4(4), 52-58 (2013).